\documentclass[prl,twocolumn,showpacs,preprintnumbers,amsmath,amssymb,tightenlines,epsfig]{revtex4}
\usepackage{graphicx}
\usepackage{color}
\newcommand{\bra}[1]{\langle\,{#1}\, |}
\newcommand{\ket}[1]{|\,{#1}\,\rangle}
\newcommand{\braket}[2]{\mbox{$\langle\,{#1}\, | \,{#2}\,\rangle$}}

\newcommand{\gstate}{\tilde{g}}
\newcommand{\hstate}{\tilde{h}}

\newcommand{\Ort}{r}

\newcommand{\Pos}{\Ort}

\newcommand{\Nrealistic}{20}
\newcommand{\Ntworealistic}{40}

\newcommand{\ssection}[1]{{\noi  \it #1:}}

\newcommand{\pdiff}[2]{\frac{\partial #1}{\partial #2}}

\newcommand{\sub}[2]{{#1}_{\mbox{\!\! \scriptsize #2}}}

\newcommand{\bv}[1]{\mathbf{ #1 }}

\def\noi{\noindent}
\def\beq{\begin{equation}}
\def\eeq{\end{equation}}

\def\CR{\nonumber\\[0.15cm]}

\newcommand{\rref}[1]{Ref.~\cite{#1}}
\newcommand{\fref}[1]{Fig.~\ref{#1}}
\newcommand{\frefp}[2]{Fig.~\ref{#1}~(#2)}

\newcommand{\eref}[1]{Eq.~(\ref{#1})}

\newcommand{\cref}[1]{chapter~\ref{#1}}

\newcommand{\Cref}[1]{Chapter~\ref{#1}}

\newcommand{\bref}[1]{(\ref{#1})}

\newcommand{\Vdipdip}{D}
\newcommand{\Ntot}{2N}

\usepackage{ulem}  
\begin{document}

\title{An optically resolvable Schr{\"o}dinger's cat from Rydberg dressed cold atom clouds}
\author{S.~M{\"o}bius, M.~Genkin, A.~Eisfeld, S.~W{\"u}ster and J.~M.~Rost}
\affiliation{Max Planck Institute for the Physics of Complex Systems, N\"othnitzer Strasse 38, 01187 Dresden, Germany}
\email{sew654@pks.mpg.de}
\begin{abstract}
In Rydberg dressed ultra-cold gases, ground state atoms inherit properties of a weakly admixed Rydberg state, such as sensitivity to long-range interactions. We show that through hyperfine-state dependent interactions, a pair of atom clouds can evolve into a spin and subsequently into a spatial Schr{\"o}dinger's cat state: The pair, containing $\Ntworealistic$ atoms in total, is in a coherent superposition of two configurations, with cloud locations separated by micrometers. The mesoscopic nature of the superposition state can be proven with absorption imaging, while the coherence can be revealed though recombination and interference of the split wave packets.
\end{abstract}

\maketitle

\ssection{Introduction} When and why mesoscopic objects begin to behave according to our classical intuition, as exemplified by Schr{\"o}dinger's famous thought-experiment~\cite{schroedinger:cat}, remains one of the fundamental questions in physics. Experimental progress to demonstrate quantum coherence in mesoscopic systems is impressive, with recent creation of superposition states of macroscopic Josephson currents~\cite{friedman:squidcat}, ten-qubit photonic cat states~\cite{Gao:hyperentangle}, six-qubit atomic hyperfine cats~\cite{Leibfried:hyperfinecat}, interference of fullerenes and even large bio-molecules~\cite{gerlich:organicmols}, superpositions of photon coherent states~\cite{takahashi:ancilla} and many more~\cite{lu:graphstates,monroe:singleatom}.

In most of these experiments the quantum mechanical superposition does not pertain to an intuitive classical observable taking common-sense values, such as the original "alive" or "dead" of Schr{\"o}dinger's cat. Instead, the superposition typically is achieved with intrinsically quantum mechanical degrees of freedom (hyperfine- or photon number states). Realizations of position-space superpositions have been limited to small delocalization lengths (several $100$nm for \rref{gerlich:organicmols}) the resolution of which requires sophisticated near-field interferometry. Here, we propose a Schr{\"o}dinger cat superposition in the relative distance of two ultra-cold atom clouds more than $10\mu$m apart. The relative distances of the two superposed configurations also differ on a micrometer scale, hence the existence of the two possible cloud configurations can be revealed with direct absorption imaging and the coherence of that superposition can be proven by interference upon recombination, 
taking Schr{\"o}dinger's cat into the optically resolvable micrometer domain.

In contrast to prior proposals with ultracold or Bose-Einstein condensed atoms (e.g.~\cite{weiss:solitoncat,streltsov:attBEC,bargill:dephasingcoll,cirac:BECsuperpos,dalvit:biggercats,dunningham:BECarrays,gordon:BECcat,hallwood:robustflow,ng:mesoscop}) we use Rydberg states, taking advantage of their inherently strong long-range interactions and short dynamical time-scales \cite{book:gallagher,molmer:review}. 
{The resulting internal forces~\cite{wuester:cradle,cenap:motion} let the system turn itself from a spin cat state into a spatial cat state.}
In addition, Rydberg systems typically allow for an accurate control of decoherence mechanisms.
\begin{figure}[htb]
\includegraphics[width=0.99\columnwidth]{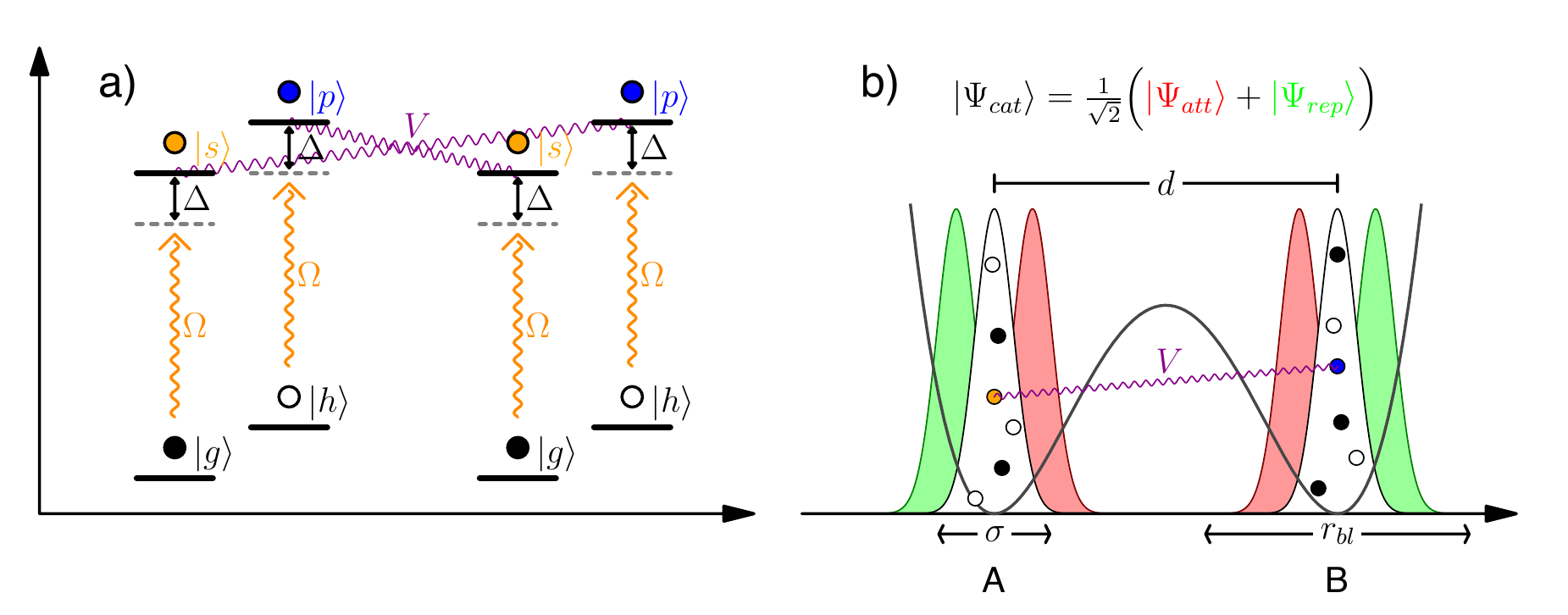}
\caption{\label{sketch}(color online) Schematic of Rydberg dressed atom clouds. (a) All atoms are in either of two hyperfine ground states $\ket{g}$, $\ket{h}$. Dressing lasers can couple one atom per cloud to either of the $\ket{s}$, $\ket{p}$ Rydberg states. The Rydberg states participate in state changing dipole-dipole interactions $\ket{sp}\leftrightarrow\ket{ps}$.
(b) Two atom clouds with width $\sigma$, separated by a distance $d$, $\sub{r}{bl}$ indicates the blockade-radius. Due to hyperfine state dependent inter-cloud forces, a suitable initial state evolves dynamically into a non-classical position space superposition state, with the pair of clouds in either the red- or green shaded configuration.
}
\end{figure}

The scheme (see \fref{sketch}) is based on a pair of atom clouds, each containing about $\Nrealistic$ alkali atoms, which can be in one of two hyperfine levels $\ket{g}$ and $\ket{h}$ of the atomic ground state.  
To induce long-range interactions between the clouds we weakly dress the states $\ket{g}$ and $\ket{h}$ with Rydberg states $\ket{s}$ and $\ket{p}$, respectively~\cite{santos:dressing,nils:supersolids,wuester:dressing,fabian:soliton}.
These are chosen such that each cloud is in the full dipole-blockade regime~\cite{jaksch:dipoleblockade}, where only a single Rydberg excitation per cloud is possible.
However, the inter-cloud distance is so large that excitations in different clouds do not block each other.
Such interactions can lead to collective relative motion of the clouds, with a repulsive or attractive character depending on the total hyperfine state. 

To realize this scheme, we first identify a suitable effective state space and Hamiltonian for our system. We then 
show how to create a hyperfine state, formally already a spin cat state \cite{opatrny:spinsqzblock,ma:spinsqzreview}, in which the two clouds evolve as a coherent superposition of attractive and repulsive dynamics.  
After a brief dwell time, single-shot absorption images would at this stage show either the green or the red configuration in \fref{sketch}.
To see the coherent character of this many-body state via interference fringes, recombination of the two configurations is finally possible with the help of an external (double-well) potential, as demonstrated in the last section.

\ssection{Ultra cold Rydberg dressing, dipole-dipole interactions and blockade}
Consider an assembly of $\Ntot$ neutral atoms of mass $M$ located at positions $\Pos_n$, restricted to one dimension and confined to a double-well atom trap. Half the atoms are localized in one of the wells, forming cloud $A$ and the rest in the other well, forming cloud $B$. Near the centres of each well at $x=\pm d/2$, the potential is approximately harmonic: $V(\Pos_n)=M \omega^2 (\Pos_n\pm d/2)^2/2$, and the atoms are initially in the Gaussian trap ground state of width $\sigma=\sqrt{\hbar/M\omega}$.
We consider four essential states in $^{87}$Rb atoms. Two of them are long lived hyperfine states $\ket{F,m_F}$, namely $\ket{g}=\ket{1,-1}$ and $\ket{h}=\ket{2,1}$ ($F$ is the total angular momentum and $m_F$ the associated magnetic quantum number). The other two essential states $\ket{\nu,l}$ are Rydberg states, designated by $\ket{s}=\ket{80,0}$ and $\ket{p}=\ket{80,1}$ ($\nu$ is the principal quantum number and $l$ the orbital angular momentum).
The Rydberg states are coupled to the ground states with Rabi frequency $\Omega$ and detuning $\Delta$, as sketched in \fref{sketch}. The coupling is off-resonant, hence $\alpha\equiv \Omega/2\Delta\ll1$. As shown in~\cite{wuester:dressing} this arrangement gives rise to effective long-range (state changing) dipole-dipole interactions of the form 
$D(r)(\ket{\gstate \hstate}\bra{\hstate\gstate}+\mbox{c.c.})$, between \emph{dressed} ground states $\ket{\gstate}\!\!\sim\!\!\ket{g} + \alpha \ket{s}$, $\ket{\hstate}\!\!\sim\!\!\ket{h} + \alpha \ket{p}$. We have $\Vdipdip(r)\approx\alpha^4 \mu^2/r^3$, where the transition dipole $\mu$ parametrizes the strength of the bare dipole-dipole interaction. Hence, we can further reduce the essential electronic state space of a single atom to $\ket{\gstate}$ and $\ket{\hstate}$, on which we build the many-body basis $\ket{\bv{k}}\equiv \ket{k_1\dots k_{\Ntot}}  \equiv\ket{{k_{1}}} \otimes \dots \otimes\ket{k_{\Ntot}}$, where $k_{j}\in \{S\}\equiv\{{\gstate},{\hstate}\}$ describes the electronic state of the atom $j$. We formulate the many-body Hamiltonian
\begin{align}
\hat{H} &= \sub{\hat{H}}{0} + \sub{\hat{H}}{int}, \:\:\:\: \sub{\hat{H}}{int} = \!\!\! \sum_{n\in A,l \in B}  \!\!\! \Vdipdip_{nl}(\bv{R})\hat{\sigma}^{(n)}_{\gstate\hstate}\hat{\sigma}^{(l)}_{\hstate\gstate} +\mbox{h.c.},
\CR
 \sub{\hat{H}}{0} &=\sum_{n=1}^{\Ntot}\left[ -\frac{\hbar^2}{2 M} \nabla^2_{\Pos_n}  +  V(\Pos_n) \right].
\label{Hamiltonian}
\end{align}
with $\hat{\sigma}^{(n)}_{kk'}=\ket{k_{n}}\bra{k_{n}'}$ where $k_n$, $k'_n$ $\in S$ in $\sub{\hat{H}}{int}$, while $\Vdipdip_{nl}(\bv{R})=\Vdipdip(|\Pos_n-\Pos_l |)$ describes the induced transition dipole-dipole interactions. Here the operator $\hat{\sigma}^{(n)}$ acts only on the Hilbertspace of atom $n$ and as unity otherwise. The vector $\bv{R}=\{r_1,\dots r_{2N}\}^T$ contains all atom co-ordinates. Note that there are no interactions between two atoms in the same cloud, since the required doubly Rydberg excited intermediate state is strongly energetically suppressed through the dipole blockade~\cite{moebius:bobbels}. The total number $N_h$ of atoms in state $\ket{\hstate}$ is used to classify the electronic states, since $N_h$ is conserved by $\sub{\hat{H}}{int}$.

Having set up our effective state space and Hamiltonian, we can construct adiabatic Born-Oppenheimer (BO) potential surfaces $U_l(\bv{R})$  
defined by $\sub{\hat{H}}{int}(\bv{R})\ket{\varphi_l(\bv{R})}=U_l(\bv{R})\ket{\varphi_l(\bv{R})}$~\cite{domke:yarkony:koeppel:CIs}. As discussed in previous work \cite{wuester:CI,moebius:cradle} the motion of atoms is determined by these BO potentials, as long as non-adiabatic effects are small, see also supplementary information \cite{sup:info}. 
We characterize Born-Oppenheimer surfaces in the vicinity of the initial configuration sketched in \fref{sketch}, with the dichotomic central many-body position $\bv{R_0}=(-d/2,\cdots,-d/2,d/2,\cdots,d/2)^T$ around which the positions of the atoms ($\bv{R}$) are randomly distributed with width $\sigma$. We choose $d=11{\mu}$m and $\sigma=0.5\mu$m. 

\begin{figure}[htb]
\includegraphics[width=0.99\columnwidth]{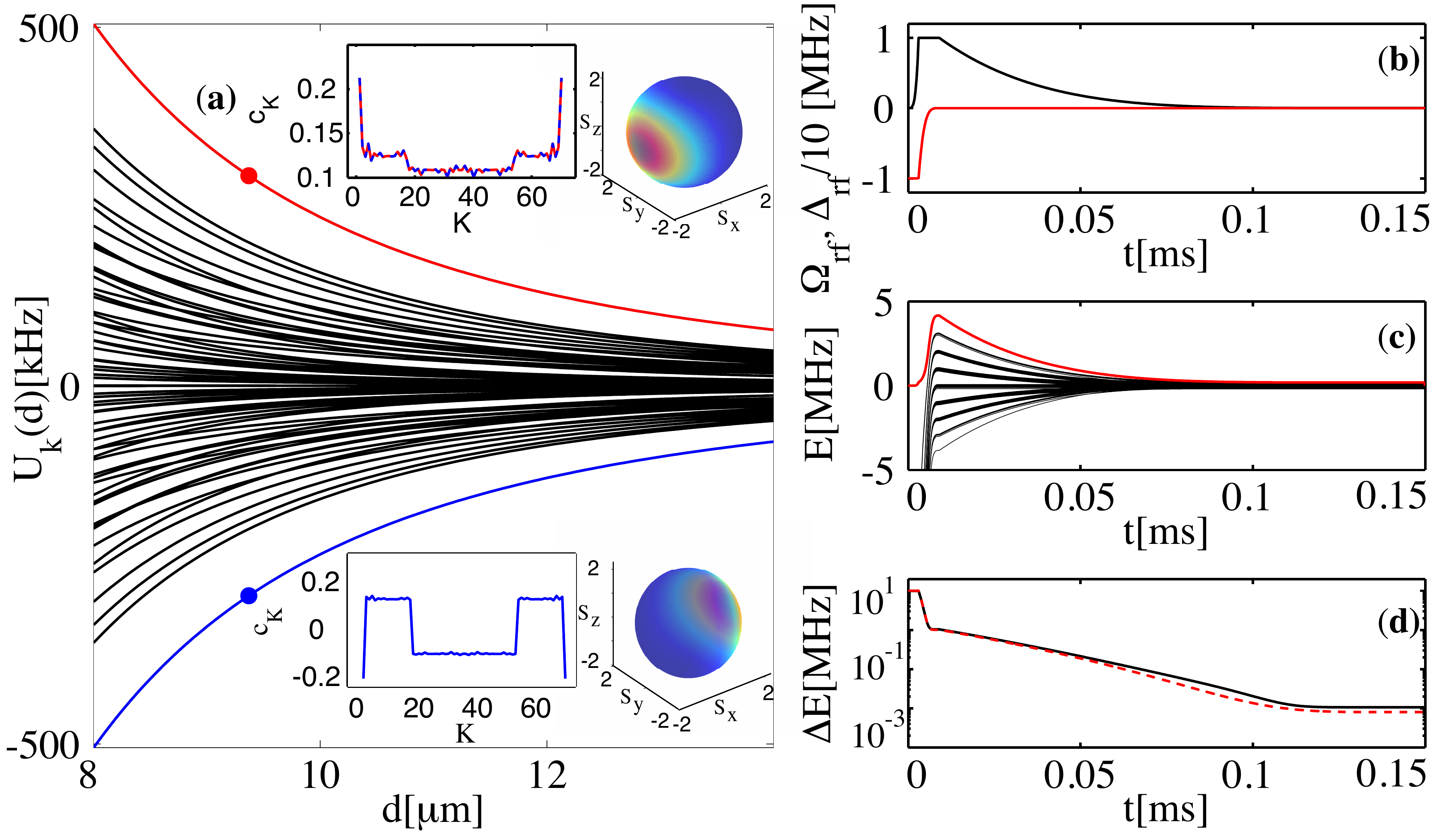}
\caption{\label{spectrum}(color online) (a) Born-Oppenheimer surfaces $U_k(\bv{R})$ as a function of $d$ for the case $\Ntot=8$, $\sigma=1\mu$m, $\alpha=0.15$ and $\mu=5760$ atomic units. Positions $\bv{R}$ have been chosen around $\bv{R_0}$ in accordance with a single realization of a Gaussian distribution (width $\sigma$). The insets show the eigenstates belonging to the two colored energies. red, top inset: Coefficients $\sub{c}{$\bv{k}$,rep}  = \braket{\bv{k}}{\sub{\Psi}{rep}}$ of the most repulsive state. blue, bottom inset: $\sub{c}{$\bv{k}$,att} = \braket{\bv{k}}{\sub{\Psi}{att}}$, most attractive.  A plot of the modulus $|\sub{c}{$\bv{k}$,att} |$ coincides with $\sub{c}{$\bv{k}$,rep} $, since the states differ only by signs of coefficients (blue dashed, top inset). The transparent spheres visualize the Q-functions of $\ket{\sub{\Psi}{rep/att}}$ in a pseudo-spin picture, in which the states resemble coherent spin states~\cite{sup:info}.
(b-d) Creation of $\ket{\sub{\Psi}{rep}}$ for $d=11{\mu}$m, using a chirped micro-wave pulse as described in the text. (b) Time dependence of micro wave Rabi frequency $\sub{\Omega}{rf}$ (black) and detuning $\sub{\Delta}{rf}/10$ (red). (c) Resulting energy spectrum of $\sub{\hat{H}}{ini}$, \eref{Hini} the red line is the state to be adiabatically followed. (d) (black) Energy gap between the two highest states of (c), compared with analytical prediction \eref{Egap} (red-dashed).}
\end{figure}

In \frefp{spectrum}{a} we show cuts through BO surfaces for states with $N_h=N$ (half the atoms in $\ket{\hstate}$) as a function of $d$. The insets show coefficients $c_\bv{k}$ of the two eigenstates $\ket{\sub{\Psi}{rep/att}}=\sum_\bv{k} \sub{c}{$\bv{k}$,rep/att} \ket{\bv{k}}$ with the largest absolute eigenvalues $\sub{U}{rep/att}(\bv{R})\approx\pm N^2 D(d)/2$. 
These states are of particular interest, since on the corresponding BO surfaces the entire clouds attract or repel, as deduced from the gradient of $\sub{U}{rep/att}$.
Consequently, after preparing the twin atom clouds in a hyperfine state $\ket{\sub{\Psi}{cat}}=(\ket{\sub{\Psi}{att}}+\ket{\sub{\Psi}{rep}})/\sqrt{2}$, one obtains a spatial superposition state as sketched in \fref{sketch} through motional dynamics. 

If our underlying basis is mapped onto a spin system \cite{sup:info}, this process can be viewed as conversion of a collective spin Schr{\"o}dinger's cat state into a spatial one. The states $\ket{\sub{\Psi}{rep/att}}$ are close to coherent spin states in this picture, as sketched in \fref{spectrum}{a}.
This conversion does not require external fields, but proceeds entirely through internal interactions within the system.
Note that the collective cloud motion in a blockade regime crucially relies on the dressed character of the interaction. For bare dipole-dipole interactions only a single atom per cloud would be accelerated~\cite{moebius:bobbels}.

Having established the fundamental mechanism behind our cat, we will outline how the initial hyperfine state $\ket{\sub{\Psi}{cat}}$ can be prepared, and then proceed to model spatial dynamics and interference.

\ssection{Initial state creation}
The first stage of assembling $\ket{\sub{\Psi}{cat}}$, starting from the simple state $\ket{\bv{\gstate}}\equiv (\ket{\bv{k}}\mbox{ with }k_n=\gstate,\:\forall n )$, is to create $\ket{\sub{\Psi}{rep}}$. This can be achieved on time scales shorter than that of atomic motion by using a micro-wave field which couples the two hyperfine-ground states so that the atom-field interaction Hamiltonian during initial state creation is~\cite{footnote:dressing_and_microwave}:
\begin{align}
\sub{\hat{H}}{ini}(\bv{R})&=\sub{\hat{H}}{int}(\bv{R})  + \sub{\hat{H}}{rf},
\CR
\sub{\hat{H}}{rf}&=\sum_{n}[\sub{\Omega}{rf}(t)\hat{\sigma}^{(n)}_{{\gstate}{\hstate}}/2 + \mbox{h.c.} -\sub{\Delta}{rf}(t)\hat{\sigma}^{(n)}_{{\hstate}{\hstate}}].
\label{Hini}
\end{align}
When we analyze the spectrum of \eref{Hini} for constant $\sub{\hat{H}}{int}$ and Rabi-frequency $\sub{\Omega}{rf}$, as a function of micro wave detuning $\sub{\Delta}{rf}$, we see that the eigenstate $\ket{\bv{\gstate}}$ at large negative detuning evolves continuously into $\ket{\sub{\Psi}{rep}}$ at $\sub{\Delta}{rf}=0$. 
This state is adiabatically followed in \frefp{spectrum}{c}, using the chirped micro-wave pulse shown in \frefp{spectrum}{b}. The pulse avoids non-adiabatic transitions since the pulse length $\sub{T}{rf}$ is long compared to the inverse energy gap ${\Delta E}^{-1}$ between the two highest energy eigenstates. The latter is well approximated by 
\begin{align}
\Delta E(t)&=\sqrt{\sub{\Delta}{rf}(t)^2 + \sub{\Omega}{rf}(t)^2}+g(N) D(d),
\label{Egap}
\end{align}
with an only weakly $N$-dependent factor $g(N)\sim 0.5$, as shown in the supplementary material \cite{sup:info}. The result \eref{Egap} simplifies the determination of realistic parameter regimes. We have numerically modelled the pulse of \frefp{spectrum}{b} for $\Ntot=8$ and found a fidelity ${\cal F}=|\braket{\sub{\Psi}{rep}}{\Psi(\sub{T}{rf})}|=0.85$ when averaging over the atomic position distribution. Our creation scheme for $\ket{\sub{\Psi}{rep}}$ closely follows the method of~\cite{pohl:crystal}. 

The second stage of initial state creation is to convert $\ket{\sub{\Psi}{rep}}$ into $\ket{\sub{\Psi}{cat}}$. We find that $\ket{\sub{\Psi}{rep}}$  and $\ket{\sub{\Psi}{att}}$ are always related as shown in \frefp{spectrum}{a}: $\ket{\sub{\Psi}{att}}$ is obtained from $\ket{\sub{\Psi}{rep}}$ by a $\pi$ phase shift to every coefficient of basis states involving an odd number $N_{hA}$, of atoms in $\ket{\tilde{h}}$ in cloud $A$. By applying this phase shift conditional on some control atom in a $(\ket{0} + \ket{1})/\sqrt{2}$ superposition we achieve our goal.
This can be realized precisely as in a recent proposal for mesoscopic Rydberg quantum computation gates \cite{mueller:EITgate}, see also \cite{sup:info}. When modelling this final step of the initial state creation sequence, we find that fidelity loss is negligible compared to the one incurred in the previous stage of creating $\ket{\sub{\Psi}{rep}}$. This situation should persist for larger $N$ \cite{mueller:EITgate}.

\begin{figure}[htb]
\includegraphics[width=0.99\columnwidth]{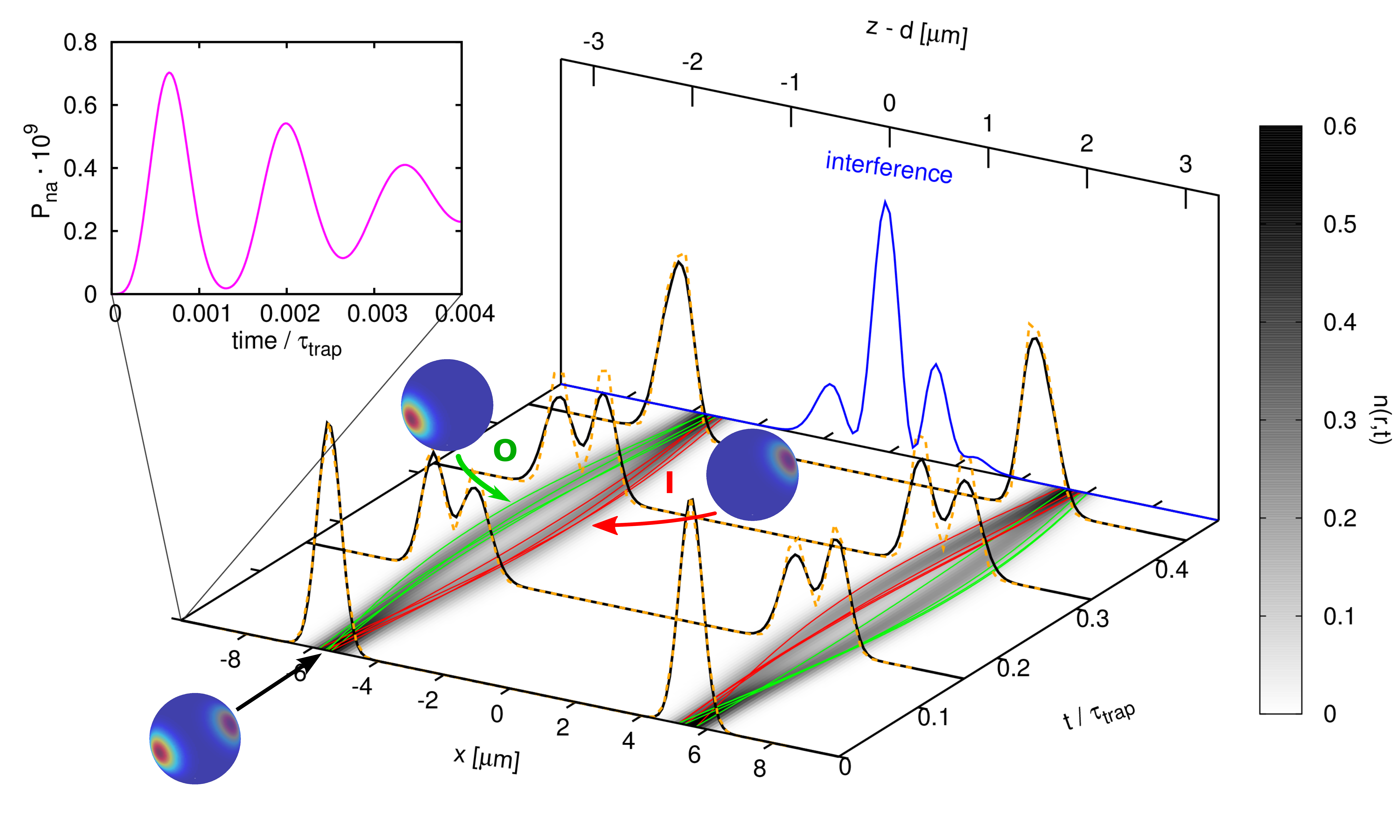}
\caption{\label{interference}(color online) Evolution of the initial hyperfine state $\ket{\sub{\Psi}{cat}}$ into a spatial superposition of cloud locations for $\Ntot=4$ and trap-frequency $\omega=(2\pi)\times400$Hz. We compare the  total atomic density $n(x,t)$ from quantum-classical simulations (grey shading, black 3D lines), to full quantum solutions (yellow dashed). The lines overlayed on the grey shading are exemplary repulsive (green) and attractive (red) quantum-classical trajectories. Transparent spheres show the conversion of a spin cat state into a spatial cat state that would underly this process for $N=20$, spin coorindate axes for spheres as in \fref{spectrum}.
Nonadiabatic population loss $P_{na}$ from the cat state during the initial acceleration phase is shown magnified in the left inset (magenta). The back panel shows the interference signal found in the relative distance distribution $\rho(z)$ at $t=\sub{\tau}{trap}/2$ (blue), note the different abscissa used. 
}
\end{figure}

\ssection{Spatial cat state creation and interference}
To turn the electronic state $\ket{\sub{\Psi}{cat}}$ prepared so far into a spatial cat, we keep the dressed interactions switched on for an acceleration period $\sub{\tau}{acc}\sim6\mu\mbox{s}$, after which they are adiabatically switched off to avoid spontaneous decay of Rydberg population. 
After mechanical evolution in the trap for a time $\sub{\tau}{trap}/4=\pi/(2\omega)$, the clouds reach their maximal displacement, where the macroscopic spatial superposition character of the quantum state can be shown with $\mu$m resolution atom detection. An absorption image would always show two inert clouds, with $50\%$ probability at either of the two configurations marked $I$ and $O$ in \fref{interference}. If instead the spatial dynamics is allowed to proceed until time $\sub{\tau}{trap}/2$ where the spatial wave function recombines and all atoms are reunited in the same hyperfine state \footnote{E.g.~by using a second chirped microwave pulse to transfer all atoms to $\ket{g}$; dressed interactions can remain off.}, we form an interference pattern, demonstrating the coherence of the superposition. 

We solve the Schr{\"o}dinger equation as in \cite{wuester:CI,sup:info,xmds:citations} to model the quantum dynamics of acceleration, splitting and recombination for $\Ntot=4$ in a plane wave basis and for $\Ntot=6$ in a Hermite-Gauss basis.
Interference fringes develop in the probability distribution $\rho(z)$ of the relative inter-cloud distance $\hat{z}=(\sum_{n\in A} \hat{R}_n-\sum_{n\in B} \hat{R}_n)/N$.
We extract  $\rho(z)$ from the many-body wave function as $\rho(z,t)=\int \overline{dR} |\Psi(\bv{R},t)|^2$, where $\int\overline{dR}$ denotes integration over all coordinates orthogonal to $z$. At $t=\sub{\tau}{trap}/2$, we find full contrast interference fringes in both cases. We thus believe that they persist also for larger atom numbers, as no new physics enters beyond $3$ atoms per cloud. Atomic densities and interference for $\Ntot=4$ are shown in \fref{interference}, which additionally includes results obtained with Tully's quantum classical algorithm~\cite{tully:hopping,wuester:cradle,moebius:cradle},
with which slightly larger atom numbers can be treated ($\sim 8 -12$).  We find that nonadiabatic effects during the acceleration phase are negligible, with a population loss of $P_{na}=1-\sub{p}{rep}=10^{-9}$ out of the target state $\ket{\sub{\Psi}{cat}}$ for the situation of \fref{interference}. For larger $N$ the situation improves further. 

%

While computational demands limit the simulations shown to $2N\leq 6,8$ we extrapolate that spatial cat states are realistic for up to $N=\Nrealistic$ for the parameters used in this article. Nonadiabatic effects during acceleration and initial state creation are under control for larger $N$. The main limitation comes from the life-time of the Rydberg states used for the dressing, since just a single decay has the potential to destroy the fragile cat state. 
However, we can choose parameters for which the probability of even a single decay is small. This is for example achieved for $d=11\mu$m, $\sigma=0.5\mu$m, $\alpha=0.15$, assuming $^{87}$Rb atoms with $\nu=80$. We use $\mu=\mu_0 \nu^2$, where $\mu_0=0.97$a.u. for $^{87}$Rb. The overall life-time of the system under dressing interactions is $\sub{\tau}{life}=\tau/(2N)$, with $\tau=\tau_0 /\alpha^2$ and $\tau_0=209.42{\mu}$s~\cite{beterov:BBR}. For our parameters $\sub{\tau}{life}=0.23$ms, larger than the time required for initial state creation ($0.15$ms) and acceleration ($6\mu$s). 

\ssection{Conclusion}
We have proposed a setup in which two cold atom clouds of about $\Nrealistic$ atoms each evolve {dynamically by internal forces} into a spatial Schr{\"o}dinger's cat state if exposed to Rydberg dipole-dipole interactions through dressing. The interactions create a state where two entire atomic clouds simultaneously are at two quantum mechanically superimposed locations, {which are macroscopically distinguishable. Hence they can be resolved by visible light.}

The internal forces that induce motion of the atomic clouds are also instrumental in creating the required intermediate hyperfine state $\ket{\sub{\Psi}{rep}}$. This state may have interesting applications by itself due to its entanglement structure between the two clouds. Finally, the hyperfine state $\ket{\sub{\Psi}{rep}}+\ket{\sub{\Psi}{att}}$ prior to any spatial dynamics realizes a collective spin Schr{\"o}dinger's cat state.

\acknowledgments
We gladly acknowledge fruitful discussions with Igor Lesanovsky, Klaus M{\o}lmer, Markus M{\"u}ller, Pierre Pillet, Thomas Pohl and Shannon Whitlock, and 
EU financial support received from the Marie Curie Initial Training Network (ITN) ÔCOHERENCE".

\section{Supplemental material}
This supplemental material provides additional details regarding the definition and use of Born-Oppenheimer surfaces, the spin structure of initial states as well as our proposal for initial state creation.

\ssection{Motion on Born-Oppenheimer surfaces}
{We insert the expansion $\ket{\Psi}=\sum_{\bv k} \phi_{\bv k}(\bv{R},t) \ket{\bv{k}}$ into the time-dependent Schr{\"o}dinger equation
$i\hbar \partial \ket{\Psi}/\partial t=\hat{H}\ket{\Psi}$ with the Hamiltonian $\hat{H}$ from Eq.~(1) of the main article. Upon projection onto the electronic basis state $\ket{\bv{k}}$ this yields a system of coupled Schr{\"o}dinger equations for the atomic motion in $\bv{R}$ and electronic dynamics}
\begin{align}
i\hbar \frac{\partial \phi_{\bv k}(\bv{R},t)}{\partial t}&= \left[ -\frac{\hbar^2}{2 M} \bv{\nabla}_{\bv{R}}^2  +  V(\bv{R}) \right] \phi_{\bv k}(\bv{R},t)
\CR
&+\sum_{\bv l} \sum_{n\in A,l \in B} \Vdipdip_{nl}(\bv{R}) \bra{\bv k} \hat{\sigma}^{(n)}_{\gstate\hstate}\hat{\sigma}^{(l)}_{\hstate\gstate} \ket{\bv  l},
\label{SE}
\end{align}
The form \bref{SE} is used in the main article.

It is often also instructive to convert to a picture using the Born-Oppenheimer separation. To this end we expand the total wave function as
\begin{align}
\ket{\Psi}&=\sum_{k} \tilde{\phi}_{\bv k}(\bv{R},t) \ket{\varphi_k(\bv{R})},
\end{align}
in terms of eigenstates of the electronic Hamiltonian
\begin{align}
\sub{\hat{H}}{int}(\bv{R})\ket{\varphi_l(\bv{R})}&=U_l(\bv{R})\ket{\varphi_l(\bv{R})}.
\label{electroniceigenstate}
\end{align}
Deriving the equation of motion from the Hamltonian using this expansion leads to the \emph{Born-Oppenheimer separated} version of the Schr{\"o}dinger equation:
\begin{align}
i\hbar \pdiff{}{t}\tilde{\phi}_l(\bv{R},t) &= \left[ -\frac{\hbar^2}{2M} \bv{\nabla}_{\bv{R}}^2  +  V(\bv{R}) + U_l(\bv{R}) \right]  \tilde{\phi}_l(\bv{R},t) 
\CR
&+ \sum_m \Lambda_{lm}(\bv{R}) \tilde{\phi}_m(\bv{R},t),
\label{BOseparatedSE}
\end{align}
where $\Lambda_{lm}(\bv{R})$ are non-adiabatic coupling terms [30]. As long as these remain small, components of the many-body wave function with different $l$ are effectively decoupled. The eigenvalues $U_l(\bv{R})$ form separate \emph{Born-Oppenheimer potential energy surfaces} that are in our context most useful to anticipate the atomic dynamics:

If the many-body wave function is localized initially near $\bv{R_0}$ in a narrow region $|\bv{R}-\bv{R_0}|<\sigma$ of the $\Ntot$ dimensional parameter-space, atom $n$ will be accelerated along the downhill gradient $F_{ln}(\bv{R_0})=-\bv{\nabla}_{r_n}U_l(\bv{R_0})$ of the energy surface $l$. Calculating these gradients for the surfaces $\sub{U}{rep}$ discussed in the main text, we find approximately $F_{{\rm rep},n}(\bv{R})\approx  \pm\frac{1}{2} N [\partial D(r)/\partial r]\big|_{r\rightarrow d}$, where the lower (upper) sign applies for an atom $n$ in cloud $A$ ($B$). $F_{{\rm att},n}$ has reverse signs.
These expressions become exact for small $\sigma$ and large $N$. The $N^2$-scaling of $\sub{U}{rep/att}$ reflects the number of interacting atom pairs, and the $N$-scaling of $F_{{\rm rep},n}$ the number of atoms excerting a force on atom $n$. Importantly, the force in those two states will induce motion of the clouds as a whole, since it is almost equally strong for all atoms. 

\ssection{Spin analogy}
After elimination of the Rydberg states our atoms are described with just two essential states, hence a mapping to coupled spin-$1/2$ particles is possible with $\ket{\downarrow} = \ket{\tilde{g}}$, $\ket{\uparrow} = \ket{\tilde{h}}$. We can then define collective spin operators for each cloud $S^{(A/B)}$
\begin{align}
\bv{S}^{(A/B)}=\sum_{n\in A/B}  \bv{s}^{(n)},
\label{spinops}
\end{align}
where the individual spin operators $ \bv{s}^{(n)}=(\sigma_x^{(n)},\sigma_y^{(n)},\sigma_z^{(n)})^T$ act on atom $n$ only, and $\sigma_j$ are Pauli matrices.

For equal interactions between all pairs of atoms from different clouds, $\Vdipdip_{nl}=D$, which is approximately realized due to $\sigma\ll d$, the interaction Hamiltonian $\sub{H}{int}$ takes the form
\begin{align}
\sub{\hat{H}}{int} = D(S_-^{(A)}S_+^{(B)} + S_+^{(A)}S_-^{(B)}),
\label{Hintspin}
\end{align}
where $S_{\pm}^{(A/B)}=S_x^{(A/B)} \pm i S_y^{(A/B)}$ are collective raising and lowering operators. Restricting ourselves to the same Hilbert space as in the main body of the paper ($2N$ atoms in both clouds together, with equal numbers in $\ket{\tilde{g}}$ and $\ket{\tilde{h}}$), we see that only collective spin states with total spin $S=N/2$ in $A/B$ and magnetic quantum numbers $m^{(A)}_s=-m^{(B)}_s=-N/2\cdots N/2$ have to be considered. In terms of the assignment of $\ket{\tilde{h}}$ states, $m^{(A)}_s = N_{hA} - N/2$, where $N_{hA}$ is the number of atoms in state $\ket{\tilde{h}}$ within cloud $A$.

The analogy to a spin model allows a spin-sphere representation of hyperfine states in our system, for which we calculate the $Q$-function defined by 
$
Q(\varphi,\theta)[\Psi] = |\braket{\varphi,\theta}{\Psi}|^2
$
for each surface element $(\varphi,\theta)$ of the sphere. These are shown in Figs.~2 and 3.

The underlying coherent spin states 
\begin{align}
\ket{\varphi,\theta}&=(1+|\eta|^2)^{-j}\sum_{m=-S}^S \sqrt{\left(\begin{array}{c} 2S \\ S+m   \end{array}\right)} \eta^{S+m}  \ket{S,m}, 
\CR
\eta &= -\tan{(\theta/2)}\exp{(-i\varphi)}
\label{CSS}
\end{align}
are defined as common in coupled spin-systems (Ref. [26]). We used $\ket{S,m} = {\cal N}\sum_{N_{hA} = m + N/2} \ket{\bv{k}}$, where ${\cal N}$ normalizes the state.

\ssection{Chirped micro-wave pulse}
In the main body of the text we have described how interaction of our system with a time-dependent micro-wave field, described by the Hamiltonian
\begin{align}
\sub{\hat{H}}{ini}(\bv{R})&=\sub{\hat{H}}{int}(\bv{R})  + \sub{\hat{H}}{rf},
\CR
\sub{\hat{H}}{rf}&=\sum_{n}[\sub{\Omega}{rf}(t)\hat{\sigma}^{(n)}_{{\gstate}{\hstate}}/2 + \mbox{h.c.} -\sub{\Delta}{rf}(t)\hat{\sigma}^{(n)}_{{\hstate}{\hstate}}],
\label{Hini}
\end{align}
can be used to adiabatically create the fully repulsive electronic state $\ket{\Psi_{rep}}$. As we are interested in fast micro-wave pulses to minimize the chance for spontaneous decay and to avoid an onset of motion, it is crucial to know the energy gap between the state that is to be adiabatically followed and other states in the spectrum.

The problem separates for the case of no interaction $\sub{\hat{H}}{int}(\bv{R})=0$, and each atom can independently be in the two eigenstates of the field Hamiltonian, which in matrix representation reads
\begin{align}
\sub{\hat{H}}{rf}^{(n)} = \left( 
\begin{array}{cc}
0 & \sub{\Omega}{rf}/2  \\
\sub{\Omega}{rf}/2 & -\sub{\Delta}{rf}
\end{array}
\right).
\label{Hn}
\end{align}
 We denote the eigenstates by $\ket{+}$ or $\ket{-}$. These have energies $E_{\pm}=(-\sub{\Delta}{rf} \pm \sqrt{\sub{\Delta}{rf}^2 + \sub{\Omega}{rf}^2})/2$. The gap between the highest energy state (all atoms in $\ket{+}$) and the state adjacent in energy (one atom in $\ket{-}$, rest in $\ket{+}$) is thus $\Delta E = E_{+} - E_{-} = \sqrt{\sub{\Delta}{rf}^2 + \sub{\Omega}{rf}^2}$.  

For the opposite case without $r.f.$ field, we can numerically solve $\sub{\hat{H}}{int}(\bv{R})$ for configurations as in Fig.~1 and find a gap $g(N) D(d)$, where the prefactor $g(N)$ is only weakly $N$-dependent and approaches $g(N)\sim 0.5$ for $N$ of the order $10$ or larger. For small $N$, e.g.,~$g(4)=1.2$.
Our analytical expression in Eq.~3 interpolates between these two limiting cases through simple addition, and is found to describe all inspected cases satisfactorily.

While the final gap for $\sub{\Omega}{rf} \rightarrow 0$ is thus roughly independent of the number of atoms, the fidelity in numerical simulations nonetheless
decreases slightly for larger $N$, due to stronger non-adiabatic couplings. It could probably be increased by pulse-shape optimization. 

\ssection{Conditional phase flip}
It was described in the main article how the maximally repulsive hyperfine state $\ket{\sub{\Psi}{rep}}$ can be obtained from a state of all atoms in $\ket{\tilde{g}}$ using a chirped microwave field. This section supplies details on the subsequent step, the transfer to the state $\ket{\sub{\Psi}{cat}}$ using the scheme of Ref.~[32].
For this we assume that dressed dipole-dipole interactions are adiabatically removed, so that $\ket{\sub{\Psi}{rep}}$ is now given in terms of \emph{bare} ground states $\ket{g}$ and $\ket{h}$.

Let there be a control atom with two internal states $\ket{0}_C$, $\ket{1}_C$, embedded in cloud $A$, which is otherwise unaffected by the creation of $\ket{\sub{\Psi}{rep}}$ so that the process described above corresponds to $\ket{0}_C\ket{\bv{\gstate}}\rightarrow \ket{0}_C\ket{\sub{\Psi}{rep}}$. The control atom could be another species of atom, or one in two hyperfine states different from $\ket{g}$, $\ket{h}$. 

Consider the following protocol:
\begin{enumerate}
\item Apply a Rabi $\pi/2$ pulse on the control atom to obtain $(\ket{0}_C + \ket{1}_C)/\sqrt{2}\ket{\sub{\Psi}{rep}}$.
\item Using the mesoscopic Rydberg quantum gate of Ref.~[32], \emph{iff the control is in $\ket{1}_C$} we now transfer all atoms in \emph{cloud $A$ only} from $\ket{h}$ into a third hyperfine state $\ket{h'}$. This creates $(\ket{0}_C \ket{\sub{\Psi}{rep}} + \ket{1}_C \ket{\sub{\Psi'}{rep}})/\sqrt{2}$, where $\ket{\sub{\Psi'}{rep}}$ is obtained from $\ket{\sub{\Psi}{rep}}$ by replacing each atom in cloud $A$ that was in state $\ket{h}$ by one in state $\ket{h'}$.
\item Apply a Rabi $2\pi$ pulse between $\ket{h'}$ and some auxiliary level to obtain a phase $(-1)$ \emph{per atom} that was in $\ket{h'}$. This has created the state $(\ket{0}_C \ket{\sub{\Psi}{rep}} + \ket{1}_C \ket{\sub{\Psi'}{att}})/\sqrt{2}$, as can be seen from Fig.~2 (a).
\item Apply the gate again to return all atoms from $\ket{h'}$ to $\ket{h}$, resulting in $(\ket{0}_C \ket{\sub{\Psi}{rep}} + \ket{1}_C  \ket{\sub{\Psi}{att}})/\sqrt{2}$, with which we have reached our goal.
\end{enumerate}
%



\begin{thebibliography}{41}
\bibitem[{foo()}]{footnote:dressing_and_microwave}
\bibinfo{note}{The microwave couples to the bare ground states. Due to the
  separate time-scales of bare dipole interactions $V\sim29$MHz, Rydberg
  dressing $\Omega\sim7.5$MHz, and microwave coupling
  $\sub{\Omega}{rf}\sim1$MHz, we expect that it adiabatically manipulates
  dressed states and write \eref{Hini} directly in terms of dressed ground
  states $\ket{\gstate}$, $\ket{\hstate}$.}
\bibitem[{xmd()}]{xmds:citations}
\bibinfo{note}{The SE and Tully's algorithm were both implemented in the
  high-level simulation language XMDS~\cite{xmds:paper,xmds:docu}.}
\end{thebibliography}
\end{document}